\documentclass[12pt]{article}
\usepackage[left=2cm, right=2cm, bottom=2cm ,top = 2cm, footskip=1cm]{geometry}
\usepackage[font=small,labelfont=bf]{caption}

\usepackage{tcolorbox}

\usepackage{setspace}
\usepackage{booktabs}
\usepackage{chngcntr}

\usepackage{titlesec}

\usepackage{enumitem}

\usepackage{hyperref}

\usepackage[authoryear,sort&compress]{natbib}
\setcitestyle{open={(},close={)}}

\usepackage{graphicx}
\usepackage{amsmath}

\usepackage[section]{placeins}

\title{Normative Modelling in Neuroimaging: \\A Practical Guide for Researchers}

\author{Nida Alyas$^{1} \textsuperscript\textdagger$, Jonathan Horsley$^{1} \textsuperscript\textdagger$, Bethany Little$^{1,2}$, \\ Peter N. Taylor$^{1,2,3}$,  Yujiang Wang$^{1,2,3}$, Karoline Leiberg$^{1*}$}
\date{}

\begin{document}

\maketitle

\begin{enumerate}
\item{CNNP Lab (www.cnnp-lab.com), School of Computing, Newcastle University, Newcastle upon Tyne, United Kingdom}
\item{Faculty of Medical Sciences, Newcastle University, Newcastle upon Tyne, United Kingdom}
\item{UCL Queen Square Institute of Neurology, Queen Square, London, United Kingdom}
\end{enumerate}

\begin{center}
\textdagger Equal contribution and first authorship

* Karoline.Leiberg@newcastle.ac.uk    
\end{center}
\newpage
\section*{Abstract}
Normative modelling is an increasingly common statistical technique in neuroimaging that estimates population-level benchmarks in brain structure. It enables the quantification of individual deviations from expected distributions whilst accounting for biological and technical covariates without requiring large, matched control groups. This makes it a powerful alternative to traditional case-control studies for identifying brain structural alterations associated with pathology.
Despite the availability of numerous modelling approaches and several toolboxes with pre-trained models, the distinct strengths and limitations of normative modelling make it difficult to determine how and when to implement them appropriately.
This review offers practical guidance and outlines statistical considerations for clinical researchers using normative modelling in neuroimaging. Through a worked example using clinical epilepsy data, we outline considerations for responsible implementation of pre-trained normative models, to support their broad and rigorous adoption in neuroimaging research.

\newpage

\section*{Introduction}

Characterising brain structure and detecting deviations can yield insight into a wide range of neurological and psychiatric disorders. Traditionally, researchers have relied on case–control designs, comparing patients against age- and sex-matched controls, for example to identify differences in brain morphology. Whilst effective for finding group-level shifts, this approach can mask the marked heterogeneity within patient populations, and is poorly suited to identifying abnormalities in individuals. Additionally, demographic mismatches between cases and controls often confound interpretation. Statistical corrections can partly mitigate these effects, but they are unstable in small cohorts and oversimplify relationships between age, sex, and brain structure.  

Normative modelling offers a more powerful alternative. Rather than contrasting cases and controls from one site, it establishes reference distributions of brain metrics such as cortical thickness, surface area, or volume from large healthy cohorts. These models predict the expected value for a given individual based on covariates such as age, sex, and scanner site, and quantify deviation against the healthy variation using z-scores or centiles. A familiar analogy is paediatric growth charts: just as a child’s height can be evaluated relative to age-matched norms \citep{Cole2012thedevelopment}, normative models assess whether a person’s brain morphology lies within or outside typical ranges. Brain morphology is a particularly compelling application area for quantitative normative modelling because, unlike standard clinical growth charts used in paediatrics, neuroimaging metrics are affected by site effects, meaning charts have to be calibrated for each new imaging site. In doing so, they enable precise, individual-level inference that is often critical in heterogeneous conditions.

Normative modelling is gaining traction across diverse applications, including epilepsy \citep{mito_towards_2025}, traumatic brain injury \citep{mitchell_normative_2025, Italinna2023}, mild cognitive impairment and dementia \citep{pinaya_using_2021, verdi_revealing_2023, verdi_personalizing_2024, Verdi2021, loreto_alzheimers_2024}, developmental psychiatry \citep{Kjelkenes2023, Holz2023}, schizophrenia \citep{lv_individual_2021, Wolfers2018}, ADHD \citep{Wolfers2020}, and autism spectrum disorder \citep{bethlehem_normative_2020, Zabihi2019, Zahabi2020}. The recent release of pre-trained models and online platforms such as Brain MoNoCle \cite{little_brain_2025}, BrainChart \cite{Bethlehem2022}, PCN Toolkit \cite{Rutherford2023}, and CentileBrain \cite{Ge2024} has made these methods more accessible to clinical researchers without extensive statistical training. By uploading processed imaging data, users can obtain deviation scores without the need to build models from scratch.

Despite these advances, clear, practical guidance on how to apply pre-trained normative models remains scarce. Researchers often face challenges such as deciding how many local controls are required for model calibration, how to deal with demographic mismatches, and whether deviation scores can be trusted without site-specific reference data. In this review, we describe the differences between normative modelling and traditional case–control approaches and provide practical recommendations for applying normative modelling in clinical neuroimaging. We illustrate each point with a worked example. Specifically, we address:
\begin{itemize}
    \item How does normative modelling differ from case-control approaches?
    \item What impact arises from demographic mismatches between cases and controls?
    \item How many healthy controls are needed to calibrate a model to a new site?
    \item Can deviation scores be reliably computed without any site-matched controls for calibration?
    \item How does the choice of normative model or platform influence results?
    \item How can we tell if a pre-trained normative model has been well calibrated to new data?
\end{itemize}
Our aim is to empower clinical researchers to adopt normative modelling with confidence, enhancing both the sensitivity and reproducibility of neuroimaging studies.

\section*{Worked example}

To demonstrate the practical impact of normative modelling in clinical research and to highlight how specific choices in using normative models influence outcomes, we present a worked example throughout this review. The example uses the IDEAS dataset \citep{taylor_imaging_2025}, consisting of individuals with medically refractory focal epilepsy and healthy controls. We focus on individuals diagnosed with left-onset mesial temporal lobe epilepsy (left mTLE), all acquired using a consistent protocol on a single scanner (n~=~23), alongside 70 matched healthy controls, or relevant subsets as indicated. Additionally, we include a case study of a 19-year-old female individual with left mTLE to illustrate individual-level deviation mapping, in the text referred to as Subject~1. Cortical thickness values were extracted using the Desikan–Killiany atlas via the FreeSurfer \textit{recon-all} pipeline. Deviations (z-scores) from the normative range were computed using pre-trained models available on the Brain MoNoCle platform \citep{little_brain_2025}.

\section*{How does normative modelling differ from case-control approaches?}


\subsection*{Case–control design}
\begin{figure}[ht!]
    \centering
    \includegraphics[scale=1]{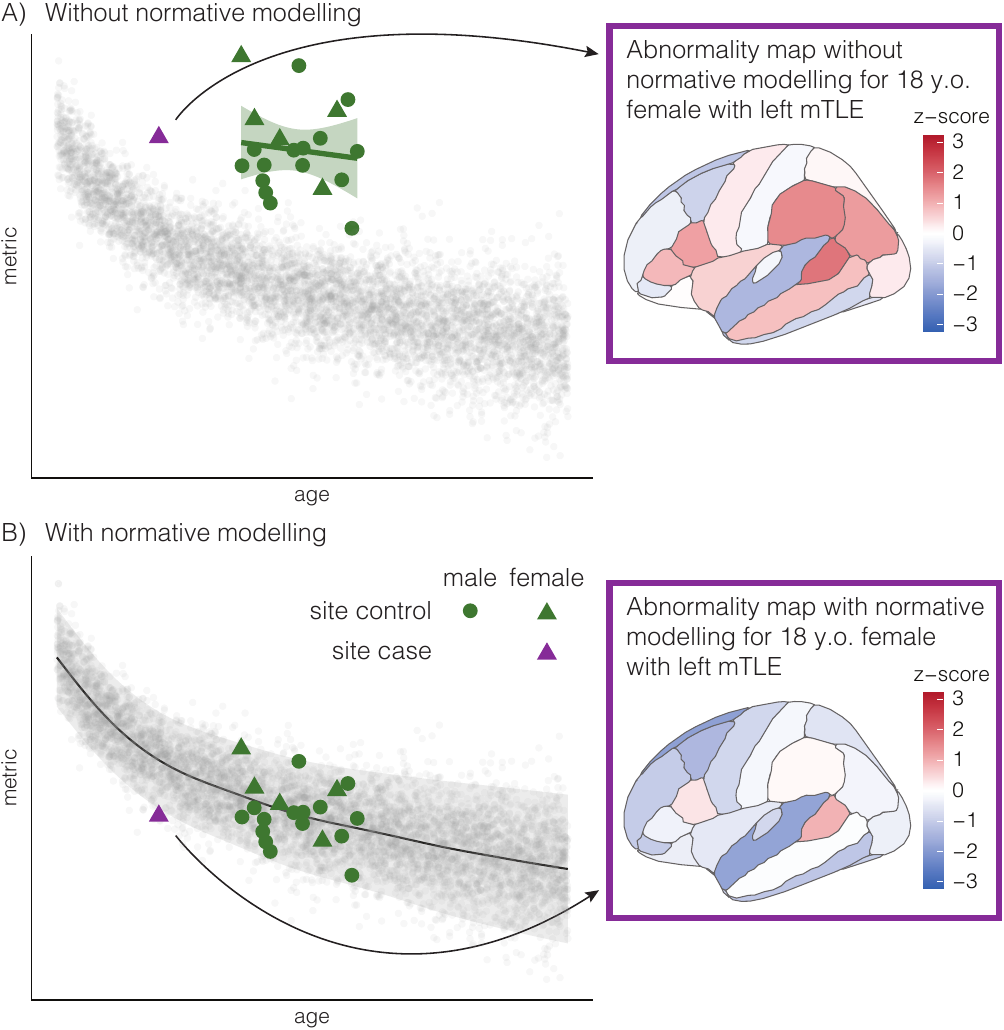}
    \caption{
        \textbf{Calculating individual abnormalities with and without normative modelling.}  
        \textbf{A:~Without normative modelling.} \textit{Left (synthetic data)}: A regional abnormality in an individual case (purple triangle) is estimated by comparing a brain structure metric to a group of site-matched controls (green points). \textit{Right}: The purple box shows the z-score abnormality map for Subject~1, derived using linear regression on age and sex based on 20~randomly sampled healthy controls. 
        \textbf{B: With normative modelling.} \textit{Left (synthetic data)}: A normative model is first trained on a large reference dataset (small, grey points). Site-specific controls (green points) are then used to calibrate the normative model to the new site, allowing the individual case (purple triangle) to be compared to the normative distribution. The model robustly accounts for covariates (e.g. age) inferred from the healthy population. \textit{Right}: The purple box shows the z-score abnormality map for Subject~1, derived using the same 20~healthy controls to calibrate the pre-trained normative model.
    }
    \label{fig:NM_motivation}
\end{figure}

In a traditional case–control study, researchers recruit demographically matched controls (often matched on age and sex) to compare against patient cohorts. This design is well suited to detect group-level differences in mean values, such as reduced hippocampal volume in Alzheimer’s disease. However, it is less effective for identifying abnormalities in individuals, where heterogeneity is high and effect sizes may be small. Consider a study that matches patients to controls at the group level: although the cohorts are balanced on average, an individual older patient may be compared to a set of younger controls, making age-related effects difficult to disentangle (Figure~\ref{fig:NM_motivation}A, Box~1). Statistical adjustments, such as regressing out age or sex, can partially mitigate this issue, but these corrections assume linear, homogeneous effects across the cohort and are often unstable in small samples \citep{kia_hierarchical_2020}, potentially introducing noise that obscures true effects \citep{little_brain_2025}. For example, Figure~\ref{fig:NM_motivation}A shows a cortical thickness z-score map for Subject~1, derived using linear regression on age and sex from a cohort of 20~healthy controls. The resulting pattern appears noisy and inconsistent, with widespread increases in cortical thickness.

\begin{tcolorbox}[colback=white,colframe=black,title=Box 1]

Practical challenges with data acquisition can make it difficult to identify abnormalities in individuals that robustly account for known, biological covariates. In Figure \ref{fig:NM_motivation}, note that:
\begin{enumerate}[topsep=0pt,itemsep=-1ex,partopsep=1ex,parsep=1ex]
    \item The site-matched controls are significantly older than the case (i.e. there is an age mismatch).
    \item The site-matched controls are mostly male but the case is female (i.e. there is a sex mismatch).
    \item There are not many site-matched controls and therefore there is uncertainty in the age-related slope. The slope estimated from the site-matched controls does not capture the slope observed in the normative population.
    \item The value of the metric in the site subjects is higher than the normative dataset and has more variance (i.e. there are systematic scanner-related offset- and variance effects).
\end{enumerate}

Normative modelling can address many of the above issues.

\end{tcolorbox}

\subsection*{Normative modelling}
Normative modelling (NM) provides a more flexible alternative. Instead of comparing cases directly to a matched control group, NM uses large healthy datasets to model the distribution of brain features as a function of covariates such as age, sex, and scanner site. These models are typically trained on multi-site cohorts spanning a broad age range, enabling them to learn both central tendencies and variability across the population. Methods commonly used include linear regression, Gaussian process regression, generalized additive models of location, scale, and shape (GAMLSS), and Bayesian hierarchical frameworks \citep{kia_hierarchical_2020, Ge2024, deBoer2024}. Unlike case–control approaches that rely on limited control samples, NM can capture complex, non-Gaussian distributions of cortical thickness, surface area, or volume, that are not linearly associated with covariates (e.g. age), better reflecting the true variability of the healthy population.

Once trained, the model predicts expected distributions of brain structure metrics for new individuals based on their covariates. Observed measures are then compared to these predictions to generate deviation scores (e.g. z-scores or centiles). These scores quantify how unusual an individual’s brain metric is, relative to peers of the same age and sex, whilst also accounting for scanner-specific effects \citep{Bayer2022}. Pre-trained models can be transferred across sites, requiring only a small calibration sample of healthy controls to adjust for scanner differences (Figure~\ref{fig:NM_motivation}B). Without such calibration, site artifacts risk being misinterpreted as pathology.

In our case study, normative modelling produces a z-score map for Subject~1 that reveals patterns of widespread cortical thickness reductions (Figure~\ref{fig:NM_motivation}B).

\subsection*{Key differences and advantages}
Normative modelling differs from case–control analysis in three important ways. First, it evaluates individuals rather than groups, allowing the detection of subject-specific deviations that would be obscured in cohort averages, whilst still allowing these individual deviations to be aggregated for subsequent group-level analyses. Second, it explicitly models biological and technical covariates, avoiding confounds that can bias group comparisons. Third, it uses large-scale training data to estimate full distributions of brain measures, enabling researchers to quantify not just mean shifts but also changes in variability or skew. These properties make NM particularly well suited for studying heterogeneous conditions such as psychiatric disorders or traumatic brain injuries, where structural abnormalities are highly individualised and case–control contrasts may yield small or inconsistent effect sizes \citep{mitchell_normative_2025}.

\section*{What impact arises from demographic mismatches between cases and controls?}


In clinical research, the reliability of outputs depends heavily on the quality and representativeness of the control cohort. Brain structure changes significantly with age and differs between males and females \citep{Frangou2021,ruigrok_meta-analysis_2014}, which is why traditional case-control studies often match participants by age and sex to reduce bias \citep{iwagami_introduction_2022}. When control groups are demographically mismatched, the resulting analysis can be misleading or confounded. Normative modelling addresses this limitation by statistically adjusting for covariates, based on patterns inferred from the normative population, rather than relying solely on site-specific controls. Site-matched controls are only used to estimate site-specific effects, such as systematic offsets, thereby enabling accurate interpretation even under suboptimal sampling conditions. To illustrate the robustness of this approach under demographic imbalance, we simulated two such scenarios in our worked example.

\begin{figure}[h!]
    \centering
    \includegraphics[scale=1]{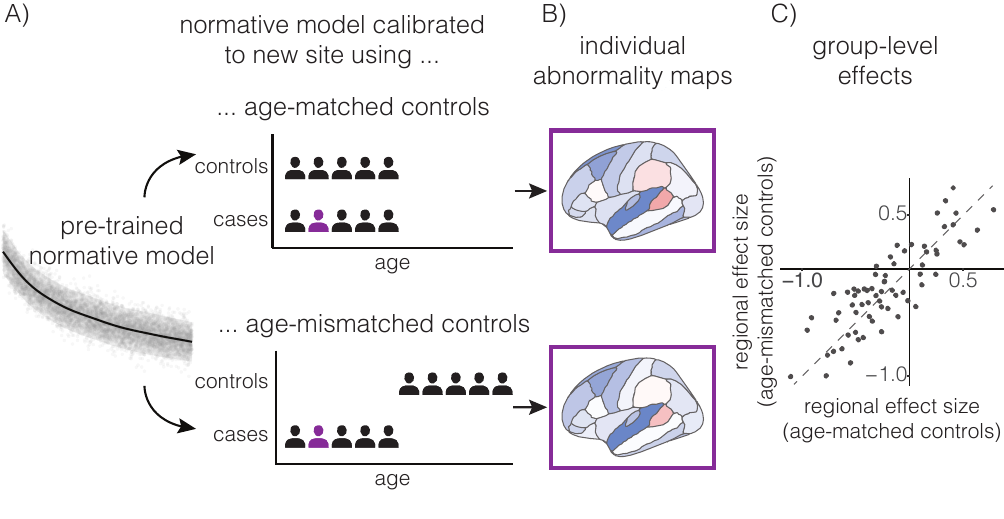}
    \caption{
        \textbf{Reproducibility of normative modelling outputs in the presence of an age mismatch between controls and cases.}
        \textbf{A:} Illustration of analysis. Pre-trained regional normative models of cortical thickness were calibrated to a new site using either age-matched or age-mismatched healthy controls.
        \textbf{B:} Calibrated models were used to derive abnormalities for individuals. The z-score abnormality map is shown here for Subject~1. 
        \textbf{C:} Regional case–control effect sizes (Cohen’s~$d$) between healthy controls and individuals with left mTLE. The plot compares group-level normative model outputs, showing regional effect sizes obtained using age-matched versus age-mismatched control groups. Each point represents a region of interest (ROI).
    }
    \label{fig:unevenage}
\end{figure}

In the first scenario, we simulated an age mismatch. A subset of individuals with left mTLE under the age of 40 was selected (n = 15). Two control groups were then constructed: one consisting of participants under 40~years old (n = 35), and another consisting of participants 40~years old and above (n = 35). Pre-trained regional normative models were calibrated to the new site using each control group separately, and used to generate individual z-score abnormality maps as well as regional effect sizes of cortical thickness comparing cases to controls (Figure~\ref{fig:unevenage}). In the case study of Subject~1, the z-score abnormality map derived using age-mismatched controls closely resembled the map obtained using age-matched controls (Figure~\ref{fig:unevenage}B). At the group level, we observed strong agreement in regional effect sizes obtained using balanced versus imbalanced control samples (Figure~\ref{fig:unevenage}C), with regional effects differing by only a small margin (mean absolute difference in regional effect size = 0.19, Pearson's correlation coefficient = 0.84). This consistency of effects despite age mismatch does not replicate in a conventional case-control analysis without normative modelling (see Supplementary Section~1).

\begin{figure}[h!]
    \centering
    \includegraphics[scale=1]{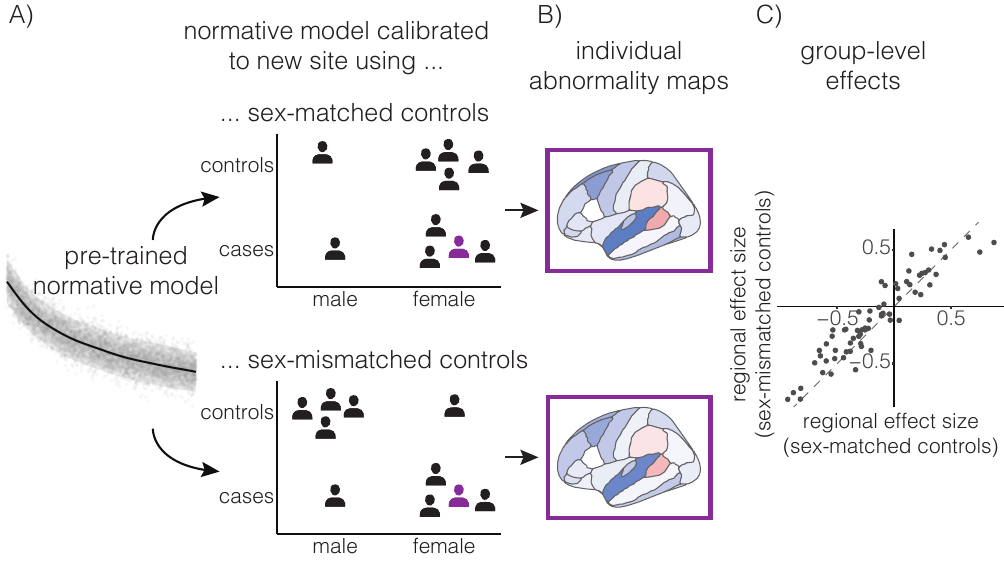}
    \caption{
        \textbf{Reliability of normative modelling outputs in the presence of a sex mismatch between controls and cases.}
        \textbf{A:} Illustration of analysis. Pre-trained regional normative models of cortical thickness were calibrated to a new site using either sex-matched or sex-mismatched healthy controls.
        \textbf{B:} Calibrated models were used to derive abnormalities for individuals. The z-score abnormality map is shown here for Subject~1. 
        \textbf{C:} Regional case–control effect sizes (Cohen’s~$d$) between healthy controls and individuals with left mTLE. The plot compares group-level normative model outputs, showing regional effect sizes obtained using sex-matched versus sex-mismatched control groups. Each point represents a region of interest (ROI).
    }
    \label{fig:unevensex}
\end{figure}

In the second simulated scenario, we introduced a sex bias. A subset of individuals with left mTLE was selected (n = 20), comprising 80\% female individuals. Again, two control groups were constructed: one with 79\% female participants (n = 33, 26 female), and another with 79\%~male participants (n = 33, 7 female).  Pre-trained regional normative models were calibrated using each control group separately, and applied to generate individual z-score abnormality maps, as well as regional cortical thickness effect sizes comparing cases to controls (Figure~\ref{fig:unevensex}). As in the age-biased simulation, the case study of Subject~1 showed that the abnormality map derived using sex-mismatched controls closely resembled the one obtained with sex-matched controls (Figure~\ref{fig:unevensex}B). At the group level, regional effect sizes were highly consistent across the two analyses (Figure~\ref{fig:unevensex}C), with minor differences in regional effects (mean absolute difference in regional effect size = 0.12, Pearson's correlation coefficient = 0.94). Similar to the age mismatch, case-control analysis using sex-matched vs sex-mismatched controls produces less consistent effects without normative modelling (Supplementary Section~2).

These examples demonstrate that normative modelling is robust in the face of demographic imbalances commonly occurring in clinical research. The general agreement between outputs derived using balanced and biased control control cohorts suggest that site-specific effects - estimated from controls - can effectively calibrate patient data, whilst covariate effects are reliably adjusted via the normative model, even when the control group is not fully representative.

Normative modelling assumes that covariate effects (age and sex) do not interact with site-specific effects, for example that a scanner-related reduction in cortical thickness measurements applies uniformly across age groups. When using pre-trained models with highly biased data, it is recommended not only to assess the overall fit of the model to the new site, but also examine whether healthy control residuals or deviation scores show age- or sex-dependent patterns (see section ``How can we tell if a pre-trained normative model has been well calibrated to new data?"). Whilst using demographically representative controls remains best practice, these simulations highlight the practical utility of normative modelling in clinical settings, where ideal control data are often difficult to obtain.

\textbf{Key takeaway: Normative models are robust to demographic mismatches; even when control cohorts are age- or sex-biased, deviation scores remain reproducible. Nonetheless, using representative controls is recommended when possible.}

\section*{How many healthy controls are needed to calibrate a model to a new site?} \label{number_healthy_controls}


Normative models are trained on large healthy cohorts to establish expected brain morphology across age and other covariates (e.g. sex, scanner site). Empirical benchmarking on over 37,000 participants shows that model convergence and stability typically require training samples of roughly 3,000 subjects \citep{Ge2024, little_brain_2025}. When applied to a new dataset, these models are calibrated using much smaller, site-specific control cohorts to infer site effects. Hierarchical Bayesian approaches further support this adaptability by using informative priors that preserve the normative baseline even with small adaptation samples \citep{kia_hierarchical_2020}. This allows patient-level evaluation without large, matched control groups at every site - a major advantage in clinical studies where healthy control numbers are often limited. However, the practical question remains: how small can the site-matched control cohort be whilst still allowing accurate site adjustment?

\begin{figure}[h]
    \centering
    \includegraphics[scale=1]{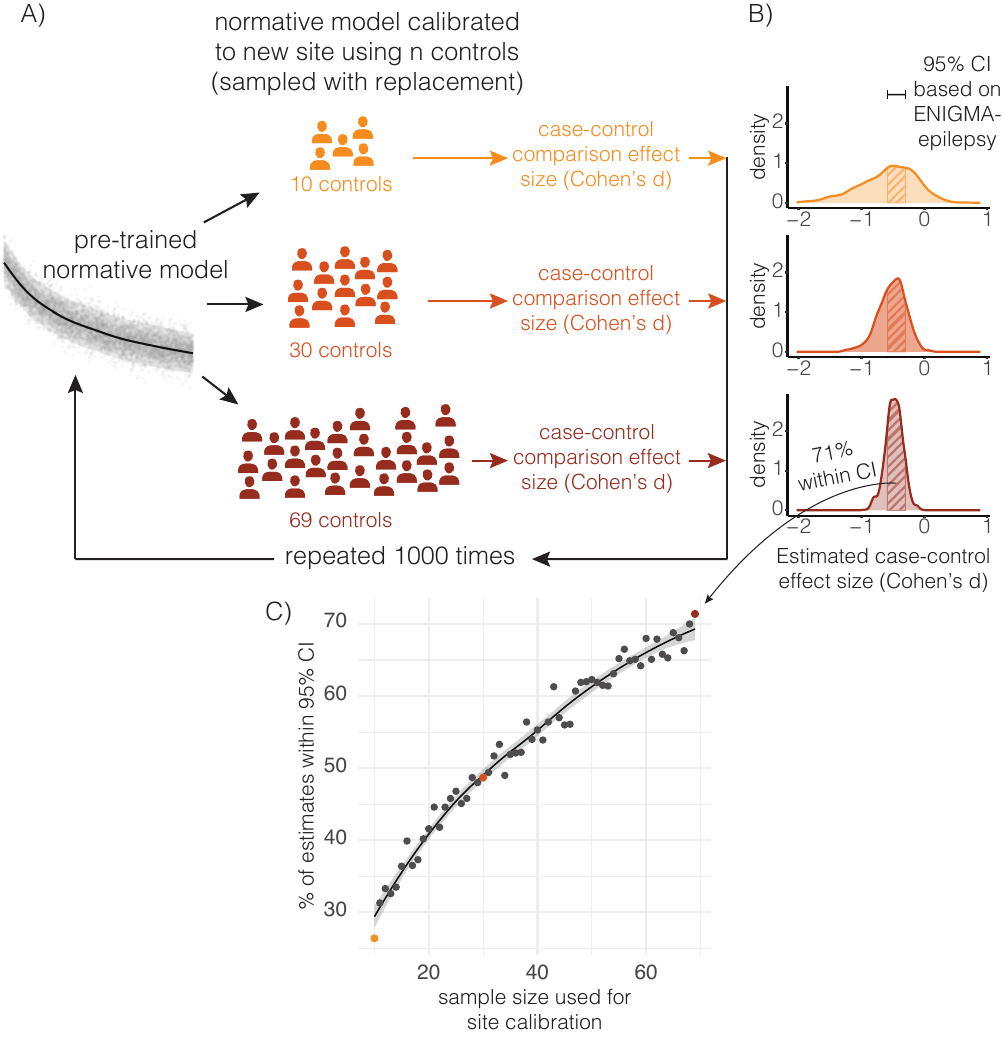}
    \caption{
        \textbf{Reliability of normative model outputs depending on healthy controls sample size used for model calibration.}  
        \textbf{A:} Illustration of analysis. A pre-trained normative model of the thickness in the left hemisphere entorhinal cortex was calibrated to a new site using varying numbers of healthy controls, sampled with replacement. The calibrated model was used to compute case–control effect sizes (Cohen’s d) between controls and individuals with left mTLE, and the procedure repeated 1,000 times for each sample size.       
        \textbf{B:} Exemplar distributions of case-control effect sizes (Cohen's d) across 1,000~repetitions for calibration with 10 (red), 30 (green), and 69 (blue) controls. Outliers over 3 standard deviations from the mean distribution value have been removed for visualisation.
        \textbf{C:} Percentage of case-control effects that fell within the 95\% confidence interval reported in \citet{Whelan2018}.
    }
    \label{fig:subsampleHC}
\end{figure}

To illustrate the impact of control sample size in our worked example, we calibrated a pre-trained normative model for cortical thickness in the left entorhinal cortex with varying numbers of healthy controls, ranging from 10 to the full set of 69 available individuals (Figure~\ref{fig:subsampleHC}A). For each sample size, we performed 1,000 repetitions, sampling with replacement to simulate drawing controls from a larger population. In each iteration, we estimated Cohen’s d as the effect size between controls and individuals with left mTLE, resulting in a distribution of effect size estimates for each sample size (Figure~\ref{fig:subsampleHC}B). To assess accuracy, we calculated the percentage of effect size estimates falling within the 95\% confidence interval for the effect of left mTLE in the entorhinal cortex reported in the ENIGMA-epilepsy study by \citet{Whelan2018} (Figure~\ref{fig:subsampleHC}C). Whilst the mean effect size remained relatively stable across control sample sizes, calibrating on smaller control groups introduced greater variability, often leading to over- or underestimation (Figure~\ref{fig:subsampleHC}C). For example, with n = 10 controls, less than 30\% of estimates fell within the confidence interval found in the ENIGMA-epilepsy study. With n = 30, consistency improved substantially, with 50\% of estimates falling within the interval.

This example demonstrates the importance of using adequately sized control groups when adjusting normative models to new sites. Although smaller samples are needed than if sex and age effects were inferred from matched controls, too few controls can distort biomarker discovery in clinical populations. Our findings using the Brain MoNoCle app suggest that normative models may perform reasonably well with a limited number of site-matched controls, producing robust estimates of site-specific mean and standard deviation. Theoretical calculations support this: for instance with as little as n = 30 controls, there is a 98\% probability of estimating the site-specific standard deviation within 30\% of its true value \citep{schillaci_estimating_2022}. However, the exact number of controls required for calibration depends on the the variability of the data at the new site, which may depend on the brain structure metric, brain region, as well as the site itself. If users of normative models have only small cohorts of healthy controls for calibration, we recommend repeated bootstrapping of controls to test for stability of outputs (see section ``How can we tell if a pre-trained normative model has been well calibrated to new data?"). Whilst small control cohorts can be sufficient, larger samples improve stability and reduce the risk of miscalibration, especially in heterogeneous datasets.

\textbf{Key takeaway: For site-specific calibration of normative models, even relatively small control cohorts may provide robust estimates, though the exact number may vary between models, metrics, regions, and site-specific variability. Very small cohorts can produce unreliable deviation scores, and larger samples remain preferable when feasible.}

\section*{Can deviation scores be reliably computed without any site-matched controls for calibration?}


In quantitative neuroimaging, technical factors such as scanner hardware, site, and acquisition protocols introduce systematic artefacts in data. Multi-site studies have repeatedly shown that these factors can confound analyses. For example, \citet{fortin_harmonization_2018} demonstrated substantial variation in cortical thickness measurements across 11 scanners, and \citet{Glocker2019} found that even after advanced preprocessing, a classifier could accurately identify the scanner used, highlighting persistent site-specific bias. Such effects can obscure pathology- or covariate-related morphological features. Accurate matching of cases with controls at the same site and using the same protocols is therefore critical for reliable deviation scores and valid inferences. In normative modelling, these site-matched controls are crucial to calibrate a trained model to a new site, since site-specific effects are estimated from controls and then applied to all data including cases.

A common challenge in clinical studies is the absence of local control data collected under the same scanning conditions as the patient cohort. Researchers may be tempted to use normative models adjusted with controls from a different site or scanner, or even just from the same scanner but different acquisition protocols, assuming statistical covariate adjustment (e.g. for age and sex) is sufficient. To illustrate how this can lead to incorrect findings, we simulated this scenario in our worked example.

\begin{figure}[h]
    \centering
    \includegraphics[scale=1]{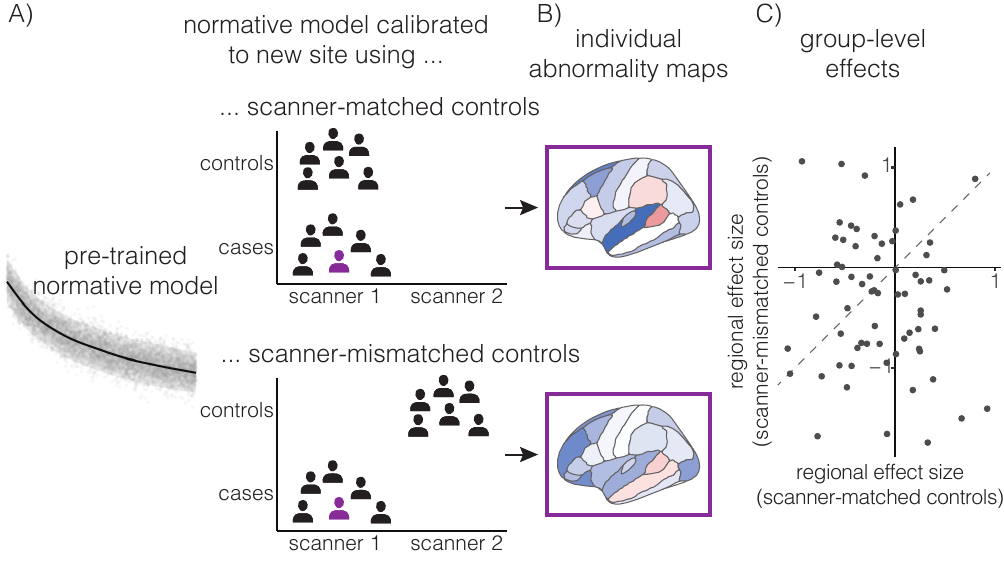}
    \caption{
        \textbf{Reliability of normative modelling outputs if controls and cases are not scanner matched.}
        \textbf{A:} Illustration of analysis. Pre-trained regional normative models of cortical thickness were calibrated to a new site using either scanner-matched or scanner-mismatched healthy controls.
        \textbf{B:} Calibrated models were used to derive abnormalities for individuals. The z-score abnormality map is shown here for Subject~1. 
        \textbf{C:} Regional case–control effect sizes (Cohen’s~$d$) between healthy controls and individuals with left mTLE. The plot compares group-level normative model outputs, showing regional effect sizes obtained using scanner-matched versus scanner-mismatched control groups. Each point represents a region of interest (ROI).
    }
    \label{fig:site_mismatch}
\end{figure}

We calibrated pre-trained models of cortical thickness using two distinct control groups: one comprising of scans (n = 30) acquired on the same scanner and using the same protocol as the cases with left mTLE in our exemplar cohort, and another using scans from a different scanner and acquisition protocol (n = 30) (Figure~\ref{fig:site_mismatch}A). Following each calibration, we generated subject-specific regional abnormality maps and computed regional effects (Cohen's d) between cases and controls. Comparison of outputs revealed substantial discrepancies between the two analyses in regional abnormality patterns for Subject~1 (Figure~\ref{fig:site_mismatch}B). Moreover, group-level effects derived using mismatched controls were uncorrelated with the findings obtained using the correctly scanner-matched controls (Figure~\ref{fig:site_mismatch}C, mean absolute difference in regional effect size = 0.66, Pearson's correlation coefficient = -0.09).

This example underscores the critical importance of scanner-matching controls and cases when applying normative models. Scanner-specific effects must be estimated for each new site. Using controls acquired under different scanning conditions can introduce systematic errors in deviation scores, producing unreliable estimates.

\textbf{Key takeaway: Deviation scores cannot be reliably computed without local controls matched by scanner and acquisition protocol. To ensure valid inference, every new dataset should include site-matched controls for model calibration.}

\section*{How does the choice of normative model or platform influence results?}


The growing availability of large-scale neuroimaging datasets, combined with advances in statistical techniques, has led to multiple pre-trained normative models of brain morphology, including Brain MoNoCle, BrainChart, PCN Toolkit, and CentileBrain. These platforms differ in model type (e.g. Generalized Additive Models for Location Scale and Shape (GAMLSS), Bayesian Linear Regression (BLR), Multivariate Fractional Polynomial Regression (MFPR), and Hierarchical Bayesian Regression (HBR), see \citet{Ge2024} for comparative benchmarking of different model types, or \citet{mitchell_normative_2025} for brief descriptions and list of advantages), underlying reference datasets, morphometric measures, and output formats. Despite their increasing adoption, the impact of model choice on downstream results is not fully established. From the model training side, the choice of algorithm and parameters can affect training efficiency and performance \citep{Ge2024}. However, here, we wanted to test if different pre-trained models differed in terms of end-user performance. 

To illustrate potential differences of pre-trained modelling platforms, we compared outputs produced for our worked example from three normative tools which incorporate pre-trained models for the Desikan-Killiany atlas: Brain MoNoCle, CentileBrain, and PCN Toolkit. We excluded BrainChart from this comparison, since its focus is on whole-brain rather than regional morphometry. We derived individual deviation scores (z-scores) from each model and  assessed agreement between z-scores across modelling platforms. We also compared regional effect sizes (Cohen’s d) between cases and controls across models.

\begin{figure}[h]
    \centering
    \includegraphics[scale=1]{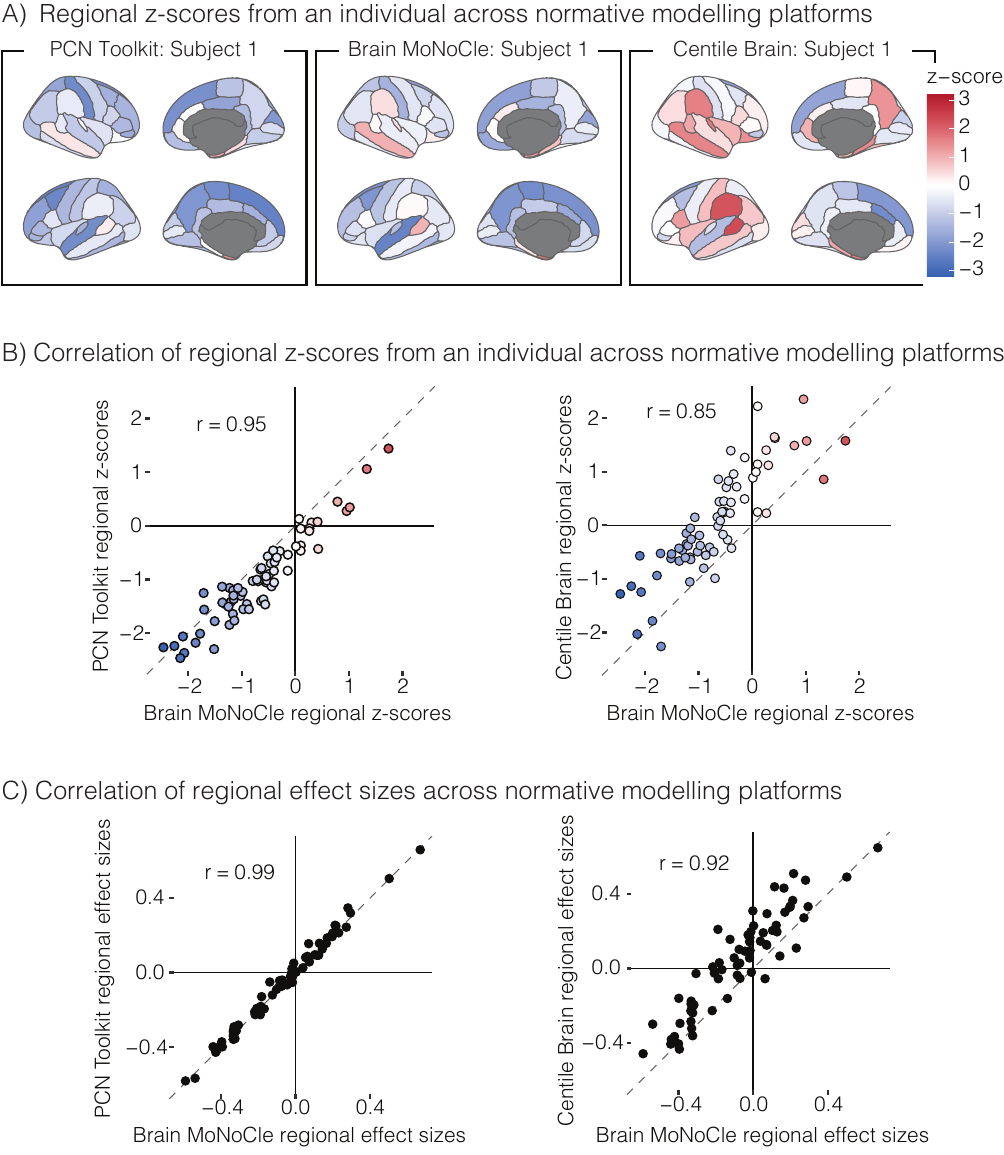}
    \caption{\textbf{Comparison of normative modelling tools in assessing cortical thickness alterations.}
    \textbf{A:} Regional z-score abnormality maps of an example subject, derived from three modelling platforms (PCN Toolkit, Brain MoNoCle, and CentileBrain).
    \textbf{B:} Correlation of regional z-score outputs between Brain MoNoCle and either PCN Toolkit (left) or Centile Brain (right) for Subject~1. Each data point represents a region of interest, coloured by z-score obtained from Brain MoNoCle. The dashed diagonal lines indicate the line of equality.
    \textbf{C:}
    Correlation of regional effect sizes between Brain MoNoCle and either PCN Toolkit (left) or Centile Brain (right) for a cohort of people with left mTLE compared to healthy controls. Effect sizes were calculated using Cohen's d. Each data point represents a region of interest. The dashed diagonal lines indicate the line of equality.
    }
    \label{fig:comparison}
\end{figure}

Despite differences in algorithms of the pre-trained models, reference datasets, and modelling frameworks, outputs were broadly consistent (Figure~\ref{fig:comparison}). Individual z-scores were in high agreement across models. 
Regional effect sizes also showed strong agreement, particularly between Brain MoNoCle and PCN Toolkit. Effect sizes from CentileBrain were systematically offset relative to Brain MoNoCle and PCN Toolkit, reflecting differences in scaling, but the relative pattern of abnormality across regions remained similar. 

These findings highlight how different pre-trained normative models provide reliable outputs at both individual- and group-levels, as they all reflect the same underlying population distribution of morphometric features. Large, representative training datasets and well-regularised models contribute to this consistency. Nonetheless, systematic offsets, such as those observed with CentileBrain, highlight that absolute values may differ even when relative patterns are preserved. Therefore, when feasible, using multiple platforms to validate key findings is advisable.

In practice, choice of platform may depend on study goals and practical considerations. These include the morphometric metrics supported (e.g. cortical thickness, volume, surface area, or other shape metrics), compatibility with specific parcellation atlases, type of outputs (e.g. z-scores vs. centiles), computational efficiency, and ease of integration into an existing analysis pipeline.

\textbf{Key takeaway: Normative modelling platform outputs generally agree, but absolute values can differ. Validate findings across tools and choose the model that best fits your metrics, atlas, and workflow.}

\section*{How can we tell if a pre-trained normative model has been well calibrated to new data?} \label{calibration_validation}

To ensure reliable outputs when applying pre-trained normative models to new datasets, it is important to verify that the model has been appropriately calibrated to the site-specific data. A well-calibrated model should yield deviation scores that reflect true biological variation rather than artefacts introduced by scanner differences or demographic factors. Calibration and model fit should be tested using control data, since assumptions about data distribution and independence of residuals may not hold for patient cohorts. When multiple independent models are calibrated (e.g. for different metrics, ROIs, or voxels), each model's fit should be assessed separately.

\begin{figure}[h!]
    \centering
    \includegraphics[scale=1]{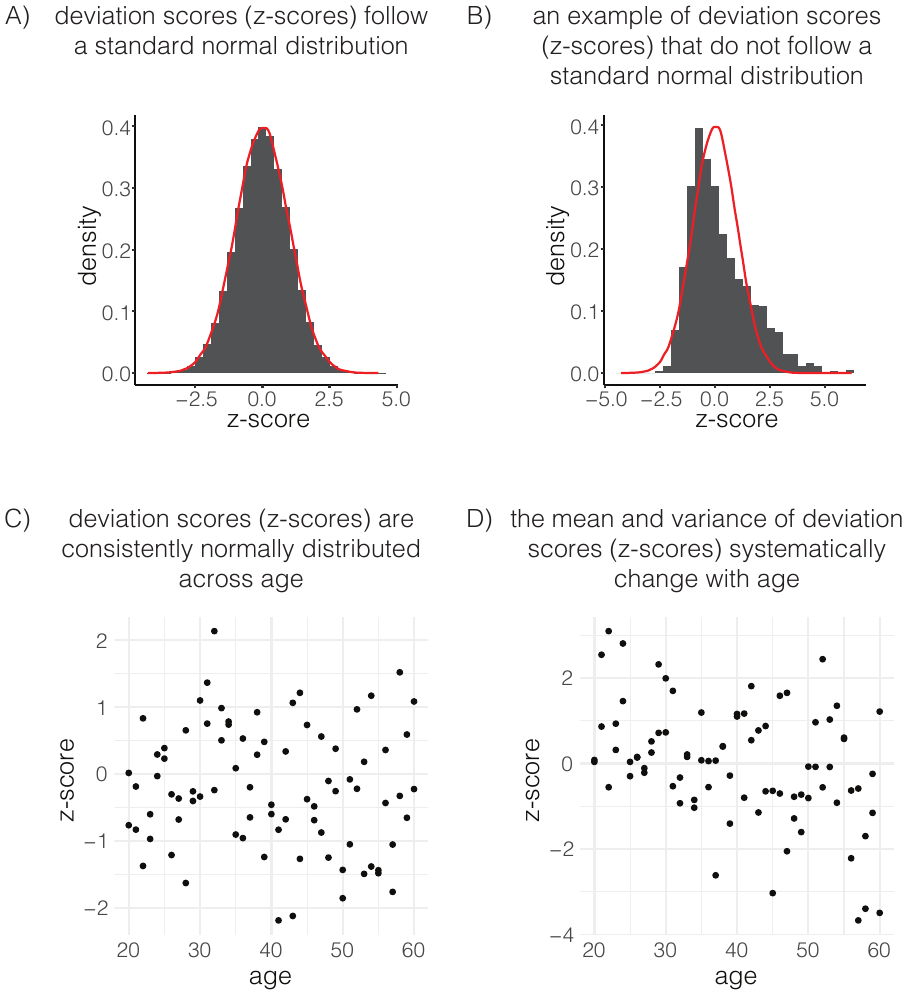}
    \caption{\textbf{Examples of assessing well-calibrated and not well-calibrated models.}
    \textbf{A:} Well-calibrated: After model calibration, z-scores of healthy controls follow a standard normal distribution (mean~=~0, variance~=~1), indicating no site-specific offset of variance remains.
    \textbf{B:} Not well-calibrated: After model calibration, z-scores of healthy controls do not follow a standard normal distribution. In this example, the distribution is left-skewed with increased variance.
    \textbf{C:} Well-calibrated: Deviation scores are independent of covariate age, indicating age effects have been sufficiently removed from new site.
    \textbf{D:} Not well-calibrated: Deviation scores depend systematically on age. In this example, z-scores decrease with age, and their variance increases with age.
    }
    \label{fig:modelFit}
\end{figure}

A practical first step is to compare the distribution of z-scores in the site-matched control group to the expected normative distribution. In a well-calibrated model, these scores should approximate a standard normal distribution (mean = 0, standard deviation = 1), which can be assessed with a Shapiro-Wilk test. Centiles should follow a uniform distribution, testable using a Kolmogorov-Smirnov or a chi-square test depending on whether centiles outputs are continuous or discrete respectively. Visual inspection can also be informative (Figure~\ref{fig:modelFit}A-B, or see e.g. \citet{dinga_normative_2021} for more examples). Deviations from the expected pattern may indicate poor calibration or uncorrected site-specific effects remaining after normative modelling.

Residuals or deviation scores can also be inspected across key covariates such as age and sex (Figure~\ref{fig:modelFit}C-D). A well-calibrated model should exhibit no systematic trends in deviation score magnitude or variance with respect to these variables (i.e. heteroskedasticity). For example, deviation scores should not change systematically with age unless this reflects a genuine biological effect. Statistical tests such as Breusch-Pagan or White test can be used to assess heteroskedasticity in residuals or deviation scores.

Stability of outputs, both at the individual- and at the group-level, can also serve as an indicator of calibration quality. This can be assessed by repeatedly refitting a model using bootstrapped controls cohorts, i.e. repeated sampling of controls with replacement, similar to the analysis in section ``How many healthy controls are needed to calibrate a model to a new site?". If individual abnormality scores or effect sizes vary substantially across different control subsamples, this may suggest that the model is sensitive to sample composition and that calibration is unstable.

Finally, where available, model fit metrics such as explained variance or reconstruction error can provide quantitative evidence of how well the model generalises to the new site. These metrics are particularly useful when working with deep normative models, where interpretability may be limited. Together, these checks offer a practical framework for evaluating calibration and ensuring that normative models yield meaningful and interpretable results when applied to new data.

\section*{Summary of recommendations}

We have examined key methodological considerations for applying normative models to clinical neuroimaging data, focusing on the impact of demographic composition of patient and control cohorts, control sample size, and scanner/protocol matching (and the influence of model choice) on results.

\subsubsection*{Demographic mismatches}
Normative models can adjust for covariates such as age and sex, making them robust to moderate demographic imbalances. Whilst demographically representative controls are ideal, deviation scores remain reliable even when the control cohort is biased.

\subsubsection*{Control sample size for site calibration}
Normative models reduce the need for large, site-matched control groups, but using very small samples for calibration can introduce high variance and unreliable deviation scores. We cannot recommend a minimum number of controls to use for stable estimates, since this may vary between models, metrics, regions, and sites, but in our experience we obtained reliable outputs when using at least 30~controls, though larger samples are generally preferred whenever feasible.

\subsubsection*{Site- and protocol-matched controls}
Accurate matching of controls and cases by scanner site and acquisition protocol is critical. Mismatched controls introduce systematic errors that can compromise deviation scores and lead to spurious findings. Site-matched controls should always be used when calibrating normative models to new datasets.

\subsubsection*{Choice of normative model or platform}
Different normative modelling tools generally produce highly consistent individual- and group-level outputs. Absolute effect sizes may differ slightly between platforms, but relative patterns of pathology are preserved. Platform selection should therefore be guided by practical considerations, including model flexibility, the structural measures of interest, atlas compatibility, output format, computational efficiency, and ease of integration.

\subsubsection*{Assess model fit and calibration}

For normative modelling to produce valid outputs, correct model calibration is essential. We therefore recommend assessing both model fit and calibration to the controls of a new sites before analysing patient abnormalities.

\subsubsection*{Overall guidance}
Normative modelling is a powerful tool for detecting brain deviations in clinical populations, offering flexibility and robustness across tools and demographic variations. Ensuring adequate control sample sizes and strict site/protocol matching maximises the reliability and clinical utility of deviation scores.

\section*{Future perspective}

Normative modelling is poised to play a central role in advancing equitable and generalisable neuroimaging research. As the field matures, future models must address current limitations in population representation, particularly the underrepresentation of ethnically diverse groups \citep{ricard_confronting_2023}. Debiasing strategies and inclusive training datasets will be essential to ensure that normative models reflect the full spectrum of human brain variation, thereby improving diagnostic accuracy and fairness across populations \citep{rutherford_which_2025}.

To achieve widespread impact, normative modelling should evolve into a standard analytical framework used by default in both research and clinical settings. Its capacity to provide individual-level insights whilst accounting for population-level variability makes it a powerful tool not only in neurology but across broader domains of biomedical imaging.

Here, we discussed normative modelling of brain morphology metrics derived from T1-weighted (T1w) MRI scans, but the same framework can be extended to other imaging and electrophysiological modalities. Different MRI sequences, such as FLAIR or diffusion MRI, may offer complementary perspectives on brain structure, whilst modalities like EEG, MEG, and fMRI provide direct measures of brain function. Extending normative modelling to these data types holds great promise for characterising individual differences across structural and functional domains. Although applications beyond T1w-MRI are still in their early stages, recent studies demonstrate its growing potential \citep{savage_dissecting_2024, lawn_normative_2024, woodhouse_multi-centre_2025, tabbal_characterizing_2025, cirstian_objective_2024, Italinna2023, villalon-reina_lifespan_2024}. As these methods mature, they may enable unified, cross-modal frameworks for robust abnormality detection within individuals.

Emerging approaches such as deep normative modelling, anomaly detection using autoencoders (AEs), and other deep learning-based techniques offer promising avenues for capturing complex, nonlinear patterns in brain data \citep{lawry_aguila_deep_2025, kumar_normative_2023}. These models can enhance sensitivity to subtle abnormalities and improve generalisation across sites and populations. However, traditional statistical models remain crucial, not only for interpretability but also for validating and guiding the development of deep learning approaches. Hybrid frameworks that integrate statistical rigour with the flexibility of deep learning may offer the best path forward.

As normative modelling continues to evolve, its integration with deep learning, privacy-preserving data sharing, and real-time clinical decision support systems will further enhance its utility. Ultimately, the future of normative modelling lies in its ability to bridge methodological innovation with clinical relevance, ensuring that neuroimaging research becomes more inclusive, reproducible, and impactful.

\section*{Conclusion}

Normative modelling is a valuable approach for individual-level brain assessment in both research and clinical settings. However, its effectiveness is highly dependent on careful methodological choices and an understanding of potential pitfalls. Normative modelling can be effective even for demographically imbalanced data, but careful assessment of model fit is required. We emphasise the importance of matching controls from the same scanner as patient cohorts to obtain reliable outputs, and recommend exercising caution in interpretation and limiting confidence in findings when only a few control subjects are available for model calibration. When observing these practical considerations and avoiding usage pitfalls, normative models are a highly reliable and powerful tool to support neuroimaging studies.

\section*{Acknowledgements}
We thank members of the Computational Neurology, Neuroscience \& Psychiatry Lab (www.cnnp-lab.com) for discussions on the analysis and manuscript. We thank Damjan Veljanoski for discussions on the manuscript. P.N.T. and Y.W. are both supported by UKRI Future Leaders Fellowships (MR/T04294X/1, MR/V026569/1).

\newpage
\bibliography{NidaNMstatisticalrecommendations}
\newpage


\setcounter{figure}{0}
\setcounter{table}{0}

\setcounter{section}{0}

\renewcommand\thesection{\Roman{section}}
\titleformat{\section}
{\normalfont\Large\bfseries}{Supplementary Text~\thesection}{1em}{}

\renewcommand\figurename{Supplementary Figure}
\renewcommand\tablename{Supplementary Table}

\renewcommand{\thefigure}{\arabic{figure}}
\renewcommand{\thetable}{\arabic{table}}
\section*{Supplementary}

\subsection*{Case-control performance using age-matched and mismatched controls}

In the main text, we showed that a pre-trained normative model maintained reproducibility when calibrated to a new site using either age-matched or age-mismatched control groups. To complement this analysis, we evaluated how a traditional case–control approach performs under the same conditions.

\begin{figure}[h!]
    \centering
    \includegraphics[scale=1]{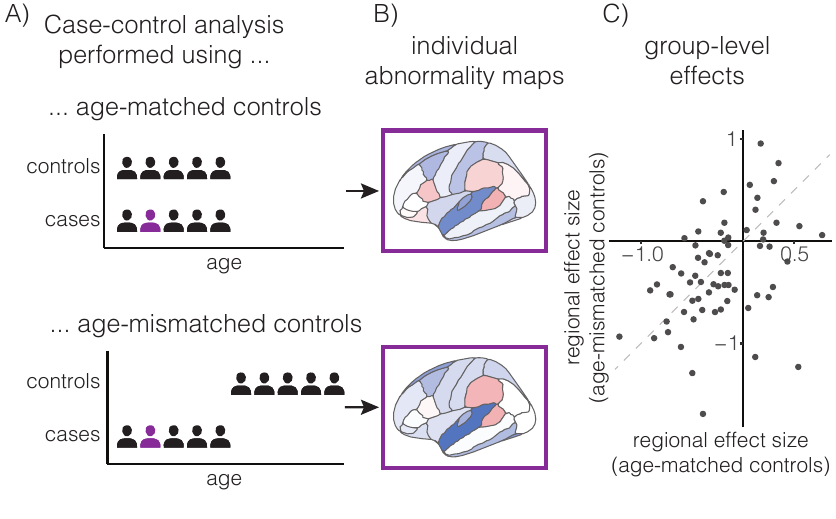}
    \caption{
        \textbf{Reproducibility of case-control outputs in the presence of an age mismatch between controls and cases.}
        \textbf{A:} Case-control analysis of cortical thickness was performed using either age-matched or age-mismatched healthy controls. Regional cortical thickness measurements were corrected for age and sex using linear regression based on healthy controls.
        \textbf{B:} Regional abnormalities in an individual were computed by z-scoring to controls. The z-score abnormality map is shown here for Subject~1. 
        \textbf{C:} Regional case–control effect sizes (Cohen’s~$d$) between healthy controls and individuals with left mTLE. The plot compares group-level case-control outputs, showing regional effect sizes obtained using age-matched versus age-mismatched control groups. Each point represents a region of interest (ROI).
    }
    \label{fig:Sup_CC_Age}
\end{figure}

At the group level, regional effect sizes obtained by comparing either to age-matched or age-mismatched control samples were moderately correlated ($r = 0.42$, Supplementary Figure~\ref{fig:Sup_CC_Age}C). This correlation was lower than that observed with the normative modelling approach ($r = 0.84$, Figure~\ref{fig:unevenage}C), indicating that case–control comparisons can be more sensitive to demographic imbalances. The result suggests that normative modelling provides a more stable estimate of effect size when controls differ in age from the target population.

\subsection*{Case-control performance using sex-matched and mismatched controls}

We next examined the effect of sex imbalance. In the main text, we demonstrated that normative modelling remained robust when calibrated with sex-matched or sex-mismatched controls. Here, we assess the corresponding behaviour of the case–control framework.

\begin{figure}[h!]
    \centering
    \includegraphics[scale=1]{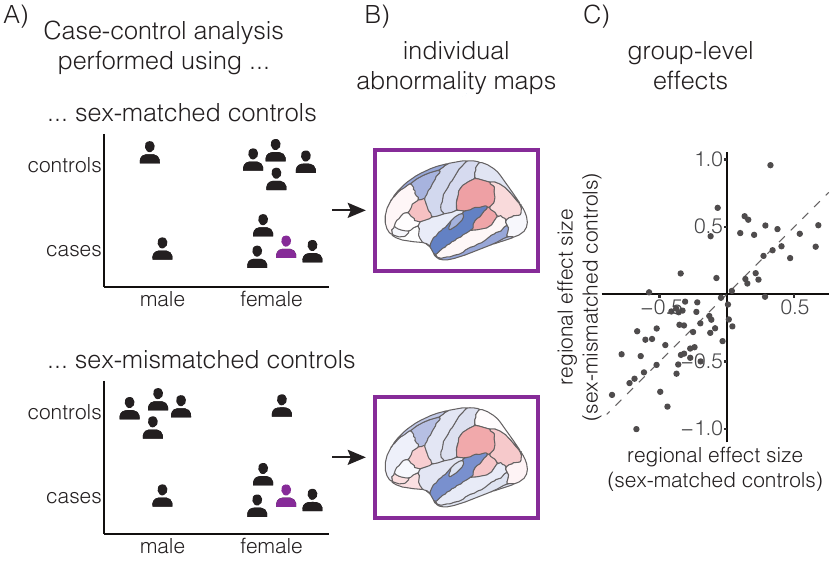}
    \caption{
        \textbf{Reproducibility of case-control outputs in the presence of an age mismatch between controls and cases.}
        \textbf{A:} Case-control analysis of cortical thickness was performed using either sex-matched or sex-mismatched healthy controls. Regional cortical thickness measurements were corrected for age and sex using linear regression based on healthy controls.
        \textbf{B:} Regional abnormalities in an individual were computed by z-scoring to controls. The z-score abnormality map is shown here for Subject~1. 
        \textbf{C:} Regional case–control effect sizes (Cohen’s~$d$) between healthy controls and individuals with left mTLE. The plot compares group-level case-control outputs, showing regional effect sizes obtained using sex-matched versus sex-mismatched control groups. Each point represents a region of interest (ROI).
    }
    \label{fig:Sup_CC_Sex}
\end{figure}

Regional effect sizes derived from sex-matched and sex-mismatched controls were strongly correlated ($r = 0.79$, Supplementary Figure~\ref{fig:Sup_CC_Sex}C), but this was again lower than the correspondence observed with the normative model ($r = 0.94$, Figure~\ref{fig:unevensex}C). The result suggests that normative modelling provides a more stable estimate of effect size when controls differ in sex from the target population.

\end{document}